\documentclass[journal]{IEEEtran}
\usepackage{mathrsfs}
\usepackage{url,times,amsmath,color,amssymb,graphicx,epsfig,psfig,cite,psfrag,subfigure,dsfont,booktabs}
\usepackage{algorithm}
\usepackage{algorithmic}
\usepackage{bm}
\usepackage{float}
\usepackage{amsmath}
\usepackage{graphicx}
\usepackage{caption}
\usepackage{subcaption} % 
\usepackage{xcolor} % 

\newtheorem{lemma}{Lemma}
\makeatletter

\newcommand{\Rmnum}[1]{\expandafter\@slowromancap\romannumeral #1@}
\makeatother

\ifCLASSINFOpdf

\else
\fi
\usepackage{cite}

\begin{document}

\title{\huge Beyond Diagonal Reconfigurable Intelligent Surface Enabled Sensing:
Cramér-Rao Bound Optimization}

\author{Xiaoqi Zhang, Liang Liu,~\IEEEmembership{Senior Member,~IEEE}, Shuowen Zhang,~\IEEEmembership{Senior Member,~IEEE}, \\Haijun Zhang,~\IEEEmembership{Fellow,~IEEE}

\thanks{
This work was supported in part by the National Key Research and Development Project of China under Grant 2022YFB2902800; in part by the National Natural Science Foundation of China under Grant 62471421. (\emph{Corresponding author: Shuowen Zhang.})

X. Zhang, L. Liu and S. Zhang are with the Department of Electrical and Electronic Engineering, The Hong Kong Polytechnic University, Hong Kong, China (e-mail: zhangxiaoqi@xs.ustb.edu.cn; liang-eie.liu@polyu.edu.hk; shuowen.zhang@polyu.edu.hk).

H. Zhang is with the Beijing Engineering and Technology Research Center for Convergence Networks and Ubiquitous Services, University of Science and Technology Beijing, Beijing 100083, China (zhanghaijun@ustb.edu.cn).
%%W. Du is with the School of Electronic and Information Engineering, Beihang University, Beijing, China, and also with the Key Laboratory of Advanced Technology, Near Space Information System (Beihang University), Ministry of Industry and Information Technology of China, Beijing, China (e-mail: wenbodu@buaa.edu.cn).
}}
\maketitle

\begin{abstract}
Recently, beyond diagonal reconfigurable intelligent surface (BD-RIS) has emerged as a more flexible solution to engineer the wireless propagation channels, thanks to its non-diagonal reflecting matrix.  Although the gain of the BD-RIS over the conventional RIS in communication has been revealed in many works, its gain in 6G sensing is still unknown. This motivates us to study the BD-RIS assisted sensing in this letter. Specifically, we derive the Cramér-Rao bound (CRB) for estimating the angle-of-arrival (AOA) from the target to the BD-RIS  under the constraint that the BD-RIS scattering matrix is unitary. To minimize the CRB, we develop an optimization scheme based on an adaptive Riemannian steepest ascent algorithm that can satisfy the non-convex unitary constraint. Numerical results demonstrate that the proposed BD-RIS-assisted target localization method achieves superior sensing performance.
\end{abstract}


\IEEEpeerreviewmaketitle

\section{Introduction}
Because integrated sensing and communication (ISAC) has been identified as a key usage scenario for 6G networks, how to provide localization service in cellular systems becomes a crucial topic. Traditional localization  techniques rely on line-of-sight (LOS) signal propagation between the target and the base station (BS) \cite{sensing}. However, in dense urban environment, the LOS path may be absent.  
In such a scenario, we can deploy a reconfigurable intelligent surface (RIS) at a proper site that is with LOS path to the target as shown in Fig. 1. Then, the BS can estimate the angle-of-arrival (AOA) of the path from the target to  RIS based on its received signals \cite{sensing1, sensing2, sensing3, duibi}. 

Recently, some novel RIS architectures such as stacked RIS, beyond diagonal RIS (BD-RIS) structure have emerged \cite{NEW1, NEW2, BDRISMAG}. Different from the conventional RIS, the circuits of various reflecting units on the BD-RIS are inter-connected such that the scattering matrix is non-diagonal.  Such a relaxation of the scattering matrix can lead to significant beamforming gain to the BD-RIS. Implementing BD-RIS hardware is challenging due to mismatches between ideal unitary constraints and practical circuit imperfections, such as complex layouts, synchronization, and calibration. Addressing these challenges requires low-complexity interconnection designs, reliable tunable components, and system-level optimization of modeling, control, and calibration \cite{duibi}.  For communication, it was shown in \cite{BDRIS4,  BDRIS2, BDRIS3} that the rate performance of a BD-RIS assisted system is significantly improved compared to the case when the conventional RIS is used. On the other hand, for localization, some heuristic metrics, e.g., localization signal-to-noise ratio (SNR), have been optimized in BD-RIS assisted systems \cite{BDRISISAC1, BDRISISAC2, BDRISISAC3}. However, in localization, Cramér-Rao bound (CRB) is an important metric, because it serves as lower bound for mean-squared error (MSE). Usually, CRB is a complicated function of the target location, especially in BD-RIS assisted systems. In this letter, we aim to design an algorithm to optimize the BD-RIS reflecting matrix for minimizing the localization CRB.  Notably, in this letter, we consider device-based sensing \cite{ sensing3, active}, which the goal is to localize the active target. The application is to localize any device that can emit radio frequency signals. For example, our technique can be utilized to localize a mobile phone, a sensor, etc. The main contributions are as follows.

%%contribution
\begin{itemize}
\item
First, this letter introduces a BD-RIS-aided localization system where the BD-RIS serves as the reference point to localize the target without LOS path to the BS, as shown in Fig. 1. To assess the target localization performance, we characterize the CRB for estimating the AOA from the target to the BD-RIS as a function of its scattering matrix.
\end{itemize}

\begin{itemize}
\item
Second, we formulate a CRB minimization problem to improve the localization performance. Under the conventional RIS aided localization system \cite{sensing1, sensing2, sensing3, duibi}, the RIS scattering matrix is a diagonal matrix with unit-power diagonal elements. Under BD-RIS aided localization system, the scattering matrix needs to be a unitary matrix because of the hardware requirement \cite{BDRIS4}. Because any diagonal scattering matrix for the conventional RIS is a feasible solution to the BD-RIS, the CRB achieved in BD-RIS aided localization system much be lower. However, optimization with a unitary matrix constraint is challenging. We tackle this issue via an adaptive Riemannian steepest ascent method \cite{RSD}.
\end{itemize}
\begin{itemize}
\item
Finally, numerical results demonstrate that the proposed BD-RIS-aided target localization scheme effectively minimizes the CRB. Compared to benchmark algorithm and traditional RIS, the proposed scheme consistently exhibits superior sensing performance.
\end{itemize}
%%

%%%
\section{System Model and Problem Formulation}
As shown in Fig. 1, this letter considers  BD-RIS aided localization in 6G ISAC network, which consists of a single-antenna active target that can emit uplink signals, a fully connected BD-RIS, and a multi-antenna BS. 
The numbers of BD-RIS reflecting elements and BS antennas are denoted by  ${N_{\rm R}}$ and ${N_{\rm BS}}$, respectively. It is assumed that the LOS path between the target and the BS is blocked, while the BD-RIS is deployed at a proper position with LOS paths to both of the target and the BS. Therefore, the signal from the target can be received by the BS merely over the target-RIS-BS path.
\begin{figure}[t]
        \centering
        \includegraphics*[width=46mm,height=32mm]{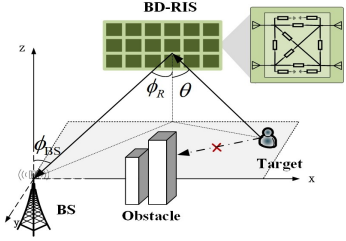}
        \caption{BD-RIS aided localization system.}
        \label{fig:1}
\end{figure}
\subsection{System Model}
Moreover, the AOA and angle-of-departure (AOD) of the path from the BD-RIS to the BS are denoted by ${{\phi  _{{\rm{BS}}}}}$ and ${{\phi  _{\rm{R}}}}$, while the AOA of the path from the target to the BD-RIS is denoted by $\theta $. 

In this letter, we assume a uniform linear array model for the BS antennas and  the RIS elements. Therefore,  given any angle $\beta$, the corresponding steering vectors of the BS and the BD-RIS towards this angle are respectively given by
\begin{equation}
{\bf{a}}_{{\rm{BS}}}^{}( \beta ) = {[1,{e^{ - j\frac{{2\pi {d_{{\rm{BS}}}}}}{\lambda }\sin \beta }},...,{e^{ - j\frac{{2\pi {d_{{\rm{BS}}}}}}{\lambda }({N_{{\rm{BS}}}} - 1)\sin \beta }}]^T},
\end{equation}
\begin{equation}
\begin{array}{l}
{\bf{a}}_{{\rm{RIS}}}^{}( \beta)\!
 = {[1,{e^{ - j\frac{{2\pi {d_{{\rm{RIS}}}}}}{\lambda }\sin \beta }},...,{e^{ - j\frac{{2\pi {d_{{\rm{RIS}}}}}}{\lambda }({N_{\rm{R}}} - 1)\sin \beta }}]^T},
\end{array}
\end{equation}
where ${{d_{{\rm{BS}}}}}$ and ${{d_{{\rm{RIS}}}}}$ denote the antenna spacing at the BS and element spacing at the BS-RIS, and $\lambda $ is the wavelength of the signal. Then, the effective channel associated with the target-RIS-BS path is given by
\begin{equation}
        {\bf{h}}( {\bf{\Phi }} ) = \alpha {\bf{a}}_{{{{\rm{BS}}}}}^{}( {{\phi  _{{\rm{BS}}}}} ){\bf{a}}_{{{\rm{RIS}}}}^H( {{\phi  _{\rm{R}}}}){\bf{\Phi }}{{\bf{a}}_{{{\rm{RIS}}}}}( {{\theta }} ),
\end{equation}
where $\alpha $ is the complex-valued channel coefficient, and ${\bf{\Phi }} \in \mathbb{C}  {^{{N_{\rm R}} \times {N_{\rm R}}}}$ is the scattering matrix of BD-RIS. 
For convenience, define ${\bf{G}} \buildrel \Delta \over = {\bf{a}}_{{{{\rm{BS}}}}}^{}( {{\phi  _{{\rm{BS}}}}} ){\bf{a}}_{{{\rm{RIS}}}}^H( {{\phi  _{\rm{R}}}} )$ as the RIS-BS channel matrix. 

Let $L$ denote the number of time slots used by the BS for AOA estimation. For the $l$-th time slot, the received signal at the BS is
\begin{equation}
  {\bf{y}}[l] =  \sqrt P {\bf{h}}( {\bf{\Phi }} ){{x}}[l] + {\bf{n}}[l], l = 1,...,L,
\end{equation}
where $P$ denotes the uplink transmit power, ${{x}}[l]$ denotes the signal transmitted by the target at time slot $l$ with $\begin{array}{*{20}{l}}
        \mathbb{E}{[|x[l]{|^2}]}
\end{array} = 1$ and ${\bf{n}}[l]\sim {\mathcal{CN}}(\bm{0}, {\sigma ^2}\rm{I})$ denotes the Gaussian noise of the BS at time slot $l$. 

By combining the signal received by the BS over all the $L$ time slots, it follows that
\begin{equation}
{\bf{Y}} = [ {{\bf{y}}[1],...,{\bf{y}}[L]}] =  \sqrt P{\bf{h}}( {\bf{\Phi }}){{\bf{x}}^H} + {\bf{N}},
\end{equation}
where ${\bf{x}} = {[x[1],...,x[L]]^H}$, and ${\bf{N}} = [{\bf{n}}[1],...,{\bf{n}}[L]]$.

\subsection{Problem Formulation}
In this letter, we aim to estimate the AOA from the target to the BD-RIS, i.e.,  $\theta$, based on the received signals $\bf{Y}$ for localization purpose. Note that besides $\theta$, the complex-valued channel coefficient $\alpha$ is an unknown and deterministic parameter. Define ${\bm{\xi }} \buildrel \Delta \over = {[ {\theta , \bm{\alpha }^T} ]^T}$ as the collection of unknown parameters that should be estimated from the received signal $\bf{Y}$, where $\bm{\alpha } \buildrel \Delta \over = {[ {\Re \{ \alpha  \},\Im \{ \alpha  \}} ]^T}$. According to \cite{FIMor}, the Fisher information matrix (FIM) for estimating ${\bm{\xi }}$ is given by
\begin{equation}
        {{\bf{F}}_{\bm{\xi }}} = \mathbb{E} {_{{\bf{Y}}|{\bm{\xi }}}}[ {\tfrac{{\partial \ln (f({\bf{Y}}|{\bm{\xi }}))}}{{\partial {\bm{\xi }}}}{{( {\tfrac{{\partial \ln (f({\bf{Y}}|{\bm{\xi }}))}}{{\partial {\bm{\xi }}}}})}^H}} ],
\end{equation}
where $ f\left( {{\bf{Y}}|{\bm{\xi }}} \right)$ is the conditional probability density function (PDF) of ${\bf{Y}}$ given ${\bm{\xi }}$. According to (5), it follows that
\begin{equation}
\begin{array}{l}
f( {{\bf{Y}}|{\bm{\xi}} }) = {( {\pi {\sigma ^2}} )^{ - {N_{{\rm{BS}}}}L}}\exp (\frac{2}{{{\sigma ^2}}}\Re \left\{ {{\rm{tr}}\left( {{{\bf{x}}^H}{{\bf{h}}^H}\left( {{\bm{\Phi}} } \right){\bf{Y}}} \right)} \right\}) \\ \qquad \quad \ \ \quad \times \exp(- \frac{1}{{{\sigma ^2}}}( {\left\| {\bf{Y}} \right\|_F^2 + \left\| {{\bf{h}}\left( {{\bm{\Phi}} } \right){\bf{x}}} \right\|_F^2} )).
\end{array}
\end{equation}
Define ${\bf{\dot h}}( {\bf{{\bf{\Phi }}}})$ as the derivative of ${\bf{h}}( {\bf{{\bf{\Phi }} }})$ with respective to $\theta$, which according to (3) is given as
\begin{equation}
        {\bf{\dot h}}\left({\bf{\Phi }}  \right) = {\rm{ - }}\tfrac{{j2\pi {d_{{\rm{RIS}}} }\alpha \cos \theta }}{\lambda }{\bf{G\Phi }}{\left[ {0,1,...,({N_{\rm{R}}} - 1)} \right]^T} \odot {{\bf{a}}_{{{\rm{RIS}}}}}\left( \theta  \right),
        \end{equation}
with $\odot$ being the Hadamard product.

By substitute (7) into (6),  the FIM matrix of  ${\bm{\xi }}$ reduces to
\begin{equation}\label{18}
\begin{array}{l}
                {{\bf{F}}_{\bm{\xi }}} = \left[ {\begin{array}{*{20}{c}}
{{{\rm{F}}_{\theta {\theta } }}}&{{{\bf{F}}_{\theta \bm{\alpha} }}}\\
{{\bf{F}}_{\theta \bm{\alpha} }^T}&{{{\bf{F}}_{\bm{\alpha} \bm{\alpha} }}}
\end{array}} \right],
\end{array}
\end{equation}
where 
\begin{equation}
        {{\rm{F}}_{\theta \theta }} = \tfrac{{2{}L}P}{{{\sigma ^2}}}{\mathop{\rm tr}\nolimits}[ {{\bf{\dot h}}( {\bf{\Phi }} ){{{\bf{\dot h}}}^H}( {\bf{\Phi }} )}],
\end{equation}
\begin{equation}
{{\bf{F}}_{\theta {\bm\alpha }}} = \tfrac{{2LP}}{{{\sigma ^2}}}\Re \{ {{ }{\rm{tr}}[ {{\bf{h}}( {\bf{\Phi }} ){{{\bf{\dot h}}}^H}( {\bf{\Phi }})} ][1,j]}\},
\end{equation}
\begin{equation}
{{\bf{F}}_{\bm{\alpha} \bm{\alpha} }} = \tfrac{{2LP}}{{{\sigma ^2}}}{\mathop{\rm tr}\nolimits} [ {{\bf{h}}( {\bf{\Phi }} ){{\bf{h}}^H}( {\bf{\Phi }})} ]{{\bf{I}}_2}.
\end{equation}

The CRB for estimating $\theta$ is given as 
\begin{equation}
{\mathop{\rm CRB_{\theta }}\nolimits} ( \bm{\Phi}) = [ {{{\bf{F}}_{\bm{\xi }}}} ]_{1,1}^{ - 1} = {[ {{{\bf{F}}_{\theta \theta }} - {{\bf{F}}_{\theta \bm{\alpha} }}{{( {{{\bf{F}}_{\bm{\alpha} \bm{\alpha} }}} )}^{ - 1}}{\bf{F}}_{\theta \bm{\alpha} }^T}]^{ - 1}}.
\end{equation}
According to (9)-(13), it follows that
\begin{equation}
        {\mathop{\rm CRB_{\theta }}\nolimits} \left( \bm{\Phi} \right)= \tfrac{{{\sigma ^2}}}{{2LP{}\left( {{\mathop{\rm tr}\nolimits} \left( {{\bf{\dot h}}\left( {\bf{\Phi  }} \right){{{\bf{\dot h}}}^H}\left( {\bf{\Phi  }} \right)} \right) - \frac{{{{\left| {{\mathop{\rm tr}\nolimits} \left( {{\bf{h}}\left( {\bf{\Phi }} \right){{{\bf{\dot h}}}^H}\left( {\bf{\Phi }} \right)} \right)} \right|}^2}}}{{{\mathop{\rm tr}\nolimits} \left( {{\bf{h}}\left( {\bf{\Phi  }} \right){{\bf{h}}^H}\left( {\bf{\Phi  }} \right)} \right)}}} \right)}}.
\end{equation}

In this letter, we aim to design the scattering matrix of the BS-RIS for minimizing the CRB of estimating $\theta$.  The problem is formulated as follows
\begin{subequations} 
\begin{align}
{\rm{}} (\rm{P}) \quad & \!\!\mathop{\min }\limits_{{\bf{\Phi }}}   \qquad {\mathop{\rm CRB_{\theta }}\nolimits} ( \bm{\Phi} ) \\
&{\rm{s.t.}}  \qquad \  {{\bf{\Phi }}^H}{\bf{\Phi }} = {\bf{I}}, 
\end{align}
\end{subequations}
where  ${{\mathop{\rm CRB_{\theta }}\nolimits}( \bm{\Phi})}$ is given in (14), and (15b) ensures that the BD-RIS scattering matrix is a unitary matrix due to the circuit limitation \cite{BDRIS4}. The unitary constraint in BD-RIS reflects the principle of energy conservation in passive and lossless networks. It ensures that total input and output power are equal and that reflected waves from different ports remain power-orthogonal, preventing inter-port energy coupling. Because a unitary matrix has orthonormal rows and columns, the system can redistribute energy across ports and spatial modes precisely and without interference.

Problem (\rm P) is non-convex optimization problem, because the CRB characterized in (14) is a non-convex function over ${\bf{\Phi }}$, and the unitary constraint is non-convex.

\section{BD-RIS Design}
%%%%
%%%%

%%%
%%%
Directly solving  Problem (P) with traditional gradient-based  methods is challenging due to the difficulty in enforcing unitary constraints at each iteration. While the unitary constraint can naturally be represented as points on a manifold, this allows for mapping the problem onto a Riemannian manifold for optimization. Such an approach leverages the manifold's geometric structure, ensuring that each iteration adheres to the unitary constraint, thus avoiding singularity issues and enabling a more efficient exploration of the solution space. In this letter, we adopt the adaptive Riemannian steepest ascent method \cite{RSD} to solve Problem (P).

For convenience, we first transform Problem (P) into the following problem
\begin{subequations} 
        \begin{align}
{\rm{}} (\rm{P'}) \quad & \!\!\mathop {\max }\limits_{\bf{\Phi }} g( {\bf{\Phi }} )\!=\!{\rm{tr}}( {{\bf{\dot h}}( {\bf{\Phi }}){{{\bf{\dot h}}}^H}( {\bf{\Phi }})} )\! - \!\tfrac{{{{| {{\rm{tr}}( {{\bf{h}}( {\bf{\Phi }} ){{{\bf{\dot h}}}^H}( {\bf{\Phi }})} )} |}^2}}}{{{\rm{tr}}( {{\bf{h}}( {\bf{\Phi }} ){{\bf{h}}^H}( {\bf{\Phi }} )} )}} \\
        &{\rm{s.t.}}  \quad\!\! 15(b). 
        \end{align}
        \end{subequations}

Next, the unitary constraint (15b) can be mapped onto a Stiefel manifold defined as 
\begin{equation}
        {\mathcal{M}}= \left\{ {{\bf{\Phi }} \in \mathbb{C}{^{{N_{\rm{R}}}\times {N_{\rm{R}}}}}:{{\bf{\Phi }}^H}{\bf{\Phi }} = {\bf{I}}} \right\}.
\end{equation}
Therefore, Problem ($\rm{P'}$) is to optimize the RIS scattering matrix on the manifold (17). 

To determine the optimal solution on the Stiefel manifold, we construct the Riemannian gradient of $g( {\bf{\Phi }} )$ and ascertain the search direction based on this gradient. The calculation of the Riemannian gradient necessitates a tangent space, which is composed of all vectors tangent to the Stiefel manifold, defined as \cite{RSD}
\begin{equation}
        {\mathcal{T}_{\bf{\Phi }}}{\mathcal{M}} =\{ {{\bf{Z}} \in\mathbb{C} {^{{N_{\rm{R}}} \times {N_{\rm{R}}}}}:{{\bf{\Phi }}^H}{\bf{Z}} + {{\bf{Z}}^H}{\bf{\Phi }} = \bf{0}}\}.
\end{equation}

Based on the tangent space in (18), we define a Riemannian metric, which is used to measure the geometric properties of the manifold, including the computation of distances, curvatures, gradients, and other complex geometric quantities.
Given any ${\bf{X}} \in {T_{\bf{\Phi }}}{\mathcal{M}}$ and ${\bf{Y}} \in {T_{\bf{\Phi }}}{\mathcal{M}}$, the Riemannian metric is defined as \cite{RSD}
\begin{equation}
        {\rho _{\bf{\Phi }}}( {{\bf{X}},{\bf{Y}}} ) =\{ {{{\left\langle {{\bf{X}},{\bf{Y}}} \right\rangle }_{\bf{\Phi }}}:{{\left\langle {{\bf{x}},{\bf{y}}} \right\rangle }_{\bf{\Phi }}} \buildrel \Delta \over = \tfrac{1}{2}\Re \left\{ {{\mathop{\rm tr}\nolimits} \left( {{\bf{X}}{{\bf{Y}}^H}} \right)} \right\}} \}.
        \end{equation}
Based on the definition of tangent space and the Riemannian metric, the Riemannian gradient of $g\left( {\bf{\Phi }} \right)$ on the Lie group of unitary matrices at ${\bf{\Phi }}$ is derived as\footnote{The derivation from the Euclidean gradient to the Riemannian gradient can be found in \cite{RSD}.}
\begin{equation}\label{18}
        \begin{array}{l}
        {\bf{\Gamma }}_{{\rm{Rie}}}^{}g( {\bf{\Phi }} ) = {\bf{\Gamma }}_{{\rm{Euc}}}^{}g\left( {\bf{\Phi }} \right) - {\bf{\Phi \Gamma }}_{{\rm{Euc}}}^Hg( {\bf{\Phi }} ){\bf{\Phi }},
        \end{array}
        \end{equation}
where ${\bf{\Gamma }}_{{\rm{Rie}}}^{}g( {\bf{\Phi }} )$ denotes the Euclidean gradient of $g( {\bf{\Phi }} )$.
According to the complex matrix gradient solution rule \cite{qiutidu}, we can derive the Euclidean gradient of $g ( {\bf{\Phi }})$ from Lemma 1.
\begin{lemma} 
Let 
\begin{equation}
        {\bf{\Omega }} = {( {{\rm{tr}}( {{\bf{h}}{{\bf{h}}^H}})})^*}{{\bf{a}}_{{\rm RIS}}}{\bf{\dot a}}_{{\rm RIS}}^H + {\rm{tr}}( {{\bf{h}}{{\bf{h}}^H}}){{\bf{\dot a}}_{{\rm RIS}}}{\bf{a}}_{{\rm RIS}}^H,
\end{equation}
the Euclidean gradient of $g ( {\bf{\Phi }})$ with respect to ${\bf{\Phi }}$ is
\begin{equation}
\begin{array}{*{20}{l}}
{\bf{\Gamma }}_{{\rm{Euc}}}^{}g( {\bf{\Phi }} )\!\!=\!\!
{  {\bf{G}}_{}^H{\bf{\dot h\dot a}}_{{\mathop{\rm RIS}\nolimits} }^H \!-\! \frac{{{\bf{G}}_{}^H{\bf{G\Phi \Omega }}}}{{{\rm{tr}}( {{\bf{h}}{{\bf{h}}^H}})}}\! + \!\frac{{{\rm{tr}}( {{\bf{h}}{{{\bf{\dot h}}}^H}}){\rm{tr}}( {{{\bf{h}}^*}{{{\bf{\dot h}}}^T}}){\bf{G}}_{}^H{\bf{ha}}_{{\mathop{\rm RIS}\nolimits} }^H}}{{{{[ {{\rm{tr}}( {{\bf{h}}{{\bf{h}}^H}} )} ]}^2}}},}
\end{array}
\end{equation}
where  ${{\bf{\dot a}}_{{\rm{RIS}}}}$ being the  derivative of ${{\bf{a}}_{{{\rm{RIS}}}}}$ about $\theta$, i.e.,
\begin{equation}
        {{\bf{\dot a}}_{{\rm{RIS}}}} = {\rm{ - }}\tfrac{{j2\pi d\cos \theta }}{\lambda }{[ {0,1,...,({N_{\rm{R}}} - 1)} ]^T} \odot {{\bf{a}}_{{\rm{RIS}}}}.
\end{equation}
\end{lemma} 
\newtheorem{proof}{Proof}
\begin{proof} 
       Please refer to Appendix A.
\end{proof} 

Next, we define the search direction on the Stiefel manifold based on the Riemannian gradient. A geodesic is the shortest local path on a manifold, enabling efficient movement analogous to straight lines in Euclidean space \cite{RSD}.

Due to the use of geodesics, the new Riemannian gradient can be obtained by performing a post-multiplication of the Riemannian gradient in (20) with ${\bf{\Phi }}^H$, i.e.,
\begin{equation}\label{18}
{\bf{\Gamma }}_{{\rm{Rie - G}}}^{}g( {\bf{\Phi }}) = {\bf{\Gamma }}_{{\rm{Euc}}}^{}g( {\bf{\Phi }} ){{\bf{\Phi }}^H} - {\bf{\Phi \Gamma }}_{{\rm{Euc}}}^Hg( {\bf{\Phi }}).
\end{equation}
In Riemannian steepest ascent method, the search or move direction of a maximization problem usually is ${\bf{\Gamma }}_{{\rm{Rie - G}}}^{}g( {\bf{\Phi }})$. Using geodesics, the rotation matrix in geodesic motion is
\begin{equation}
        {{\bf{R}}^{}} ={{\rm {exp}}({  \mu {\bf{\Gamma }} _{\rm Rie - \rm G}^{}g ( {\bf{\Phi }} )})},
\end{equation}
where $\mu  > 0$ denotes the step size. Thus the update of ${\bf{\Phi }}$ in Riemannian manifold can be write as 
\begin{equation}\label{18}
        \begin{array}{l}
{{{\bf{\Phi }}}^{( {t + 1})}} = {{\bf{R}}^{( t )}} {{\bf{\Phi }}^{( t )}}\textcolor{blue}{,}
\end{array}
\end{equation}
where $t$ denotes the iteration. 

In fact, a fix step size often struggles to balance local and global optimality. To increase the probability of escaping local optimal, we adopt adaptive step sizes that adjust dynamically based on optimization conditions. 

The design process of the BD-RIS matrix optimization for Problem ($\rm{P'}$) is summarized as Algorithm 1.  Given the non-convexity of the problem, Algorithm 1 uses a random unitary matrix to initialize the scattering matrix, improving the likelihood of better convergence.

Algorithm 1 achieves low complexity through the use of geodesics and adaptive step sizes. Owing to the properties of the geodesic exponential map, doubling the step size only requires squaring the matrix, avoiding the need for a new matrix exponential \cite{RSD}. The adaptive step size also has lower complexity compared to the classic Armijo rule. In each iteration, the step size is evaluated for suitability. If it is too small, it is doubled; if too large, it is halved. This avoids recomputing a new matrix. To reduce the high computational complexity of matrix exponential operations, efficient approximation techniques such as truncated Taylor series and diagonal Padé approximations can be used to further alleviate the computational burden. The computational complexity of computing the Euclidean gradient is ${\cal O}( {{N_{\rm BS}}N_{\rm R}^2 + N_{\rm R}^3} )$, the complexity of computing the Riemannian gradient is  ${\cal O}( {N_{\rm R}^3})$, and the complexity of matrix multiplication along the geodesic path is  ${\cal O}( {N_{\rm R}^3})$. Let ${I_{{\rm{total}}}}$ and ${I_{{\rm{\mu}}}}$ denote the number of iterations for Riemannian steepest ascent  optimization and step size adjustment, respectively. Then, the overall computational complexity of Algorithm 1 is ${\cal O}( {{I_{\rm{total}}}( {{N_{\rm BS}}N_R^2 + {I_\mu }N_{\rm R}^3})})$.

Additionally, our proposed algorithm demonstrates numerical stability. Given the recursive nature of the algorithm, usually, there will be a deviation between the new update value and the unitary constraint. Since the rotation operator maps updates back to the manifold at each iteration, errors do not accumulate and the unitary constraint is automatically preserved. Specifically, the rotation update operation (26) preserves unitary invariance, ensuring that the updated $\bm{\Phi}$ consistently satisfies the unitary constraint, thereby maintaining numerical stability. 
\setcounter{algorithm}{0}
\renewcommand\baselinestretch{1}
\begin{algorithm}[t]\vspace{0pt}
\renewcommand{\thealgorithm}{1}
\caption{BD-RIS Design Algorithm for Problem ($\rm{P'}$)}
\small
\begin{algorithmic}[1]
\STATE    Initialize $t$=0, set  ${\bf{\Phi }}^{(0)}$, tolerance threshold $\varepsilon$.
\REPEAT 
\STATE    Update  ${\bf{\Gamma }}_{{\rm{Euc}}}^{\left( t \right)}g\left( {\bf{\Phi }} \right)$ by (22).
\STATE    Update  ${\bf{\Gamma }}_{{\rm{Rie - G}}}^{\left( t \right)}g\left( {\bf{\Phi }} \right)$  by (24).
\STATE    Calculate $\eta {\rm{ = }}\rho _{\bf{\Phi }}^{( t )}( {{\bf{\Gamma }}_{{\rm{Rie - G}}}^{( t }g( {\bf{\Phi }}),{\bf{\Gamma }}_{{\rm{Rie - G}}}^tg( {\bf{\Phi }} )} ) $ by (19).
\STATE    Calculate ${{\bf{R}}^{(t)}}$ by (25).
\WHILE   {$g ({{\bf{R}}^{(t)}}{\bf{ \Phi }}^{( {t })}) - g ({{\bf{\Phi }}^{(t)}}) < \frac{\mu }{2}{ \eta}$}
\STATE    $\mu=\frac{\mu }{2} $.
\ENDWHILE
\WHILE    {$g ({{\bf{R}}^{(t)}}{{\bf{R}}^{(t)}}{\bf{\Phi }}^{\left( {t } \right)})\! - g ({{\bf{\Phi }}^{(t)}}) \ge \mu { \eta} $}
\STATE    $\mu =2\mu $. 
\ENDWHILE
\STATE    Update ${{\bf{R}}^{(t)}}$ by (25).
\STATE    Update ${{\bf{\Phi }}^{\left( t+1 \right)}}$ by (26).
\STATE  $t=t+1 $.
\UNTIL  $| {g( {{{\bf{\Phi }}^{(t)}}} ) - g( {{{\bf{\Phi }}^{(t-1)}}} )} |/g( {{{\bf{\Phi }}^{(t-1)}}} ) \le \varepsilon$.
\end{algorithmic}
\end{algorithm}

In the Stiefel manifold defined in (17), each column is a unit vector, which implies that the elements of the matrix are bounded in magnitude. Moreover, it forms a closed set in Euclidean space. Since it is both bounded and closed, the space is compact. On such a structure, any continuous function is guaranteed to attain its optimum. Therefore, the proposed algorithm is guaranteed to converge to the global optimum.

%%%
%%%%

%%%%
%%%%
\section{Numerical Results}

The numerical parameters are as follows. The noise power of sensing signal is -120 dBm, with 256 time slots.  The BS has 8 transmit antennas, while the BD-RIS contains between 8 and 64 elements. The path-loss exponents for both the target-RIS and RIS-BS links are set to 2.0. The positions of the BD-RIS and BS are [0 m, 20 m] and [-10 m, 0 m], respectively.

In this letter, we compare the proposed algorithm's efficiency and superiority against three benchmark schemes.
\begin{itemize}
        \item \textbf{Benchmark Scheme I}: The Riemannian conjugate gradient algorithm, as presented in \cite{BDRIS4}, is employed.
\end{itemize}
\begin{itemize}
        \item \textbf{Benchmark Scheme II}: The traditional RIS structure is optimized based on the SDR method proposed in \cite{duibi}.
\end{itemize}
\begin{itemize}
        \item \textbf{Benchmark Scheme III}: The scattering matrix of BD-RIS is randomly generated under the unitary constraint.
\end{itemize}

In Fig. 2, the BD-RIS is configured with 64 elements. Fig. 2 presents the convergence curves of the proposed scheme in comparison to three benchmark schemes. The results indicate that the proposed algorithm outperforms the benchmarks in terms of localization CRB performance. Additionally, it converges faster than Benchmark Scheme I.
 \begin{figure}[t]
        \centering
        \includegraphics*[width=48mm,height=36mm]{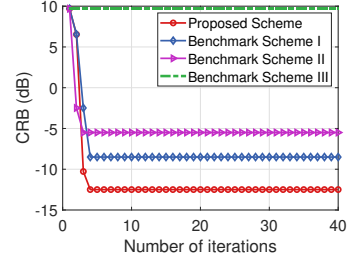}
        \caption{Convergence of different scheme.}
        \label{fig:1}
\end{figure}
 \begin{figure}[t]
        \centering
        \includegraphics*[width=48mm,height=36mm]{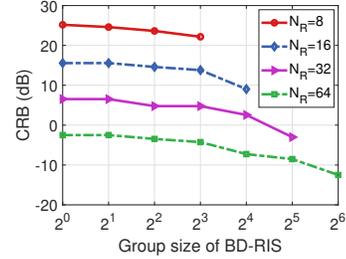}
        \caption{CRB versus BD-RIS group size.}
        \label{fig:1}
\end{figure}

Fig. 3 presents that increasing the BD-RIS group size improves AoA estimation accuracy. It shows that a full connected BD-RIS achieves the same CRB performance using 32 elements as a single connected RIS with 64, due to increased degrees of freedom enabling flexible beamforming. 
This also allows fewer active elements in target localization. 
\begin{figure*}
	\centering
                \begin{minipage}[b]{0.31\textwidth}
                        \centering
                        \includegraphics[width=48mm,height=36mm]{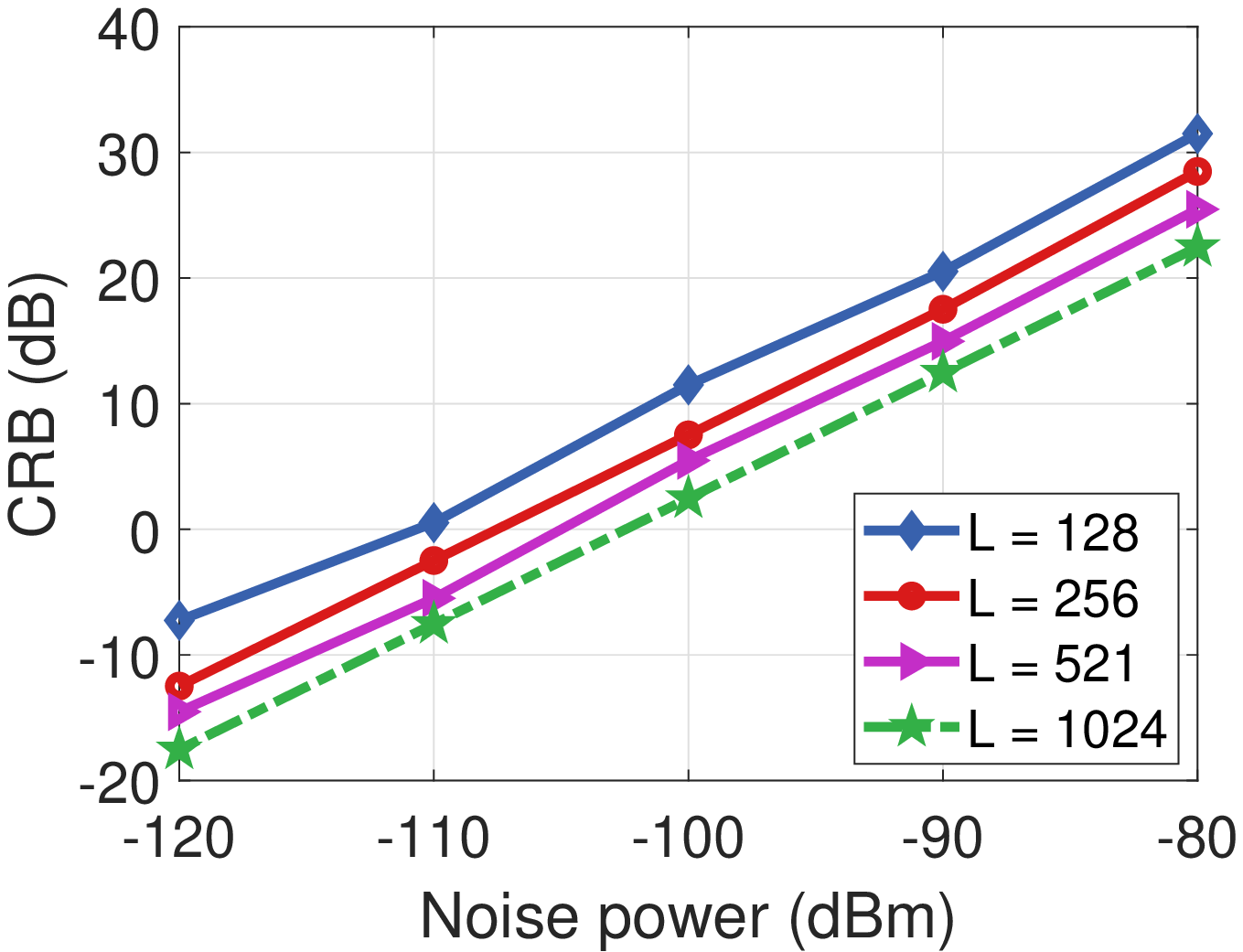}
                        \caption{CRB versus noise power.}
                    \end{minipage}
                    \hfill
                    \begin{minipage}[b]{0.31\textwidth}
                        \centering
                        \includegraphics[width=48mm,height=36mm]{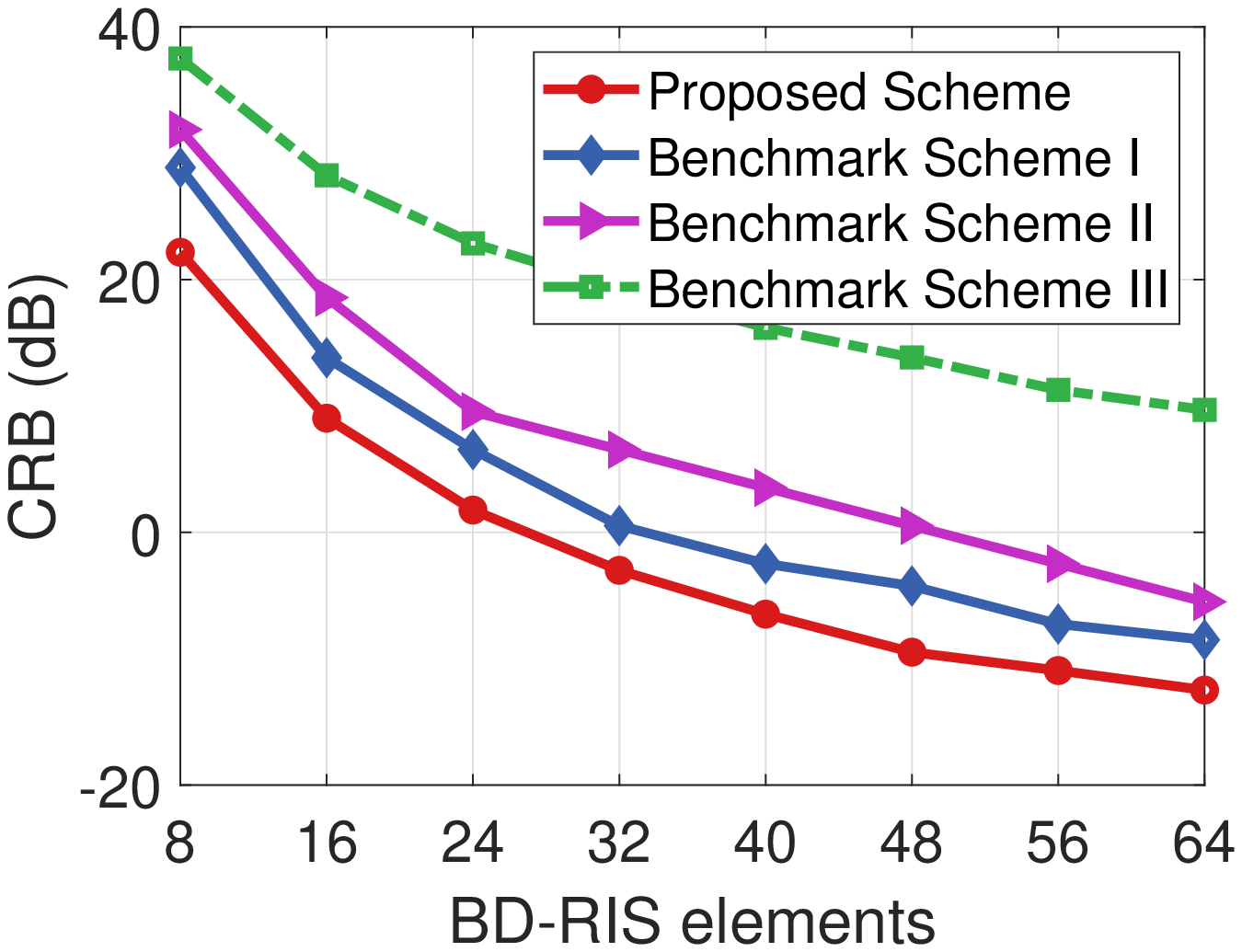}
                        \caption{CRB versus BD-RIS elements.}
                    \end{minipage}
                    \hfill
                    \begin{minipage}[b]{0.31\textwidth}
                        \centering
                        \includegraphics[width=48mm,height=36mm]{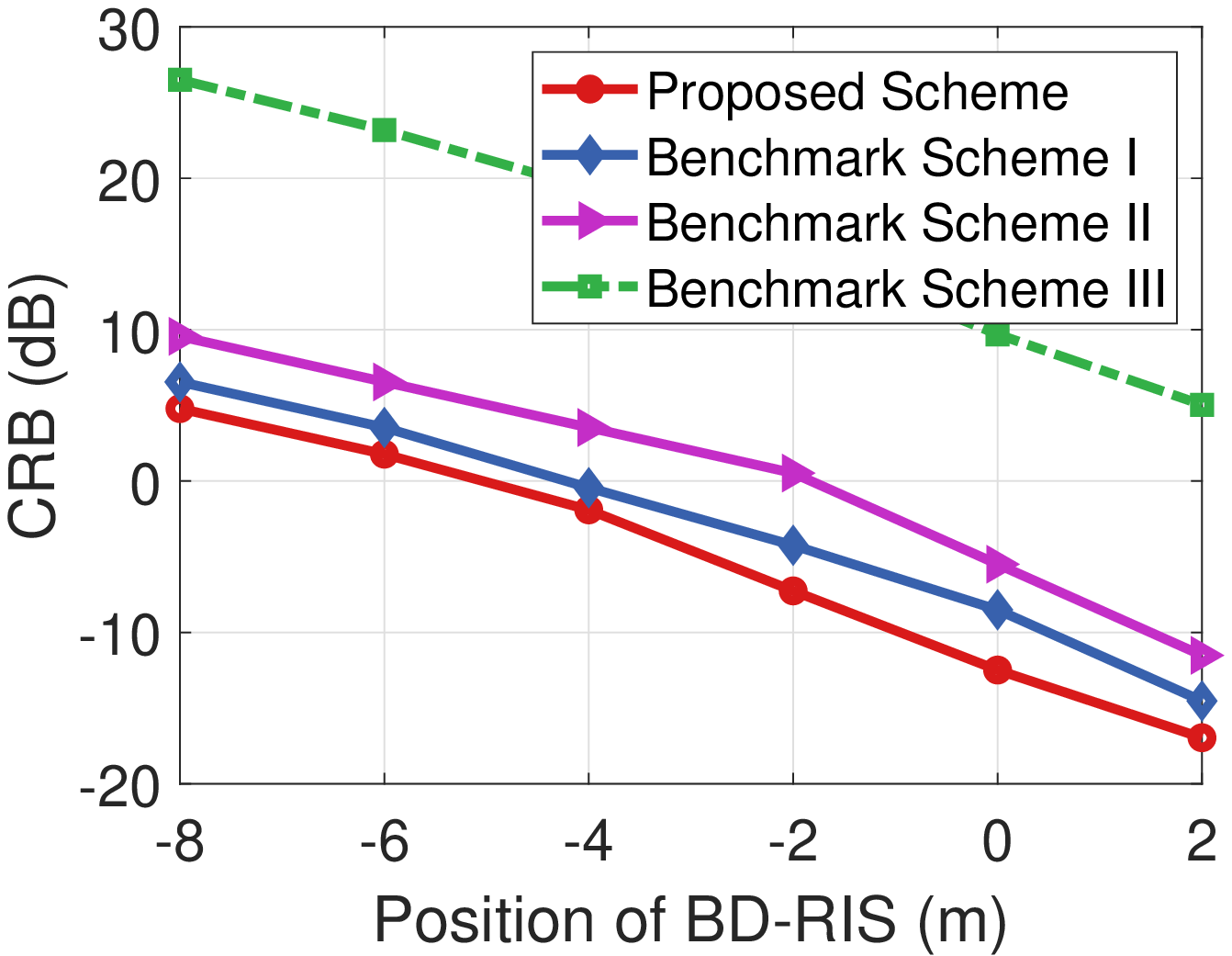}
                       \caption{CRB versus position of BD-RIS.}
                    \end{minipage}
\end{figure*}

Fig.4 illustrates the impact of the number of time slots and noise levels on the CRB. It is observed that CRB decreases with the number of time slots and increases with noise level.

Fig. 5 shows that all four schemes achieve better CRB performance with more BD-RIS elements.
The proposed scheme outperforms Benchmark Schemes I and II, and compared to Scheme III, the scattering matrix optimization proves essential for minimizing the CRB. This highlights the importance of both the beyond-diagonal design and matrix optimization for improving target sensing. 

Fig. 6 illustrates how CRB performance varies with BD-RIS position. The position here refers to the vertical coordinate of BD-RIS is fixed at 20 m, and the horizontal coordinate  varies along the x-axis in Fig. 6. The number of BD-RIS elements is 64. Assuming the BS position is fixed, the simulation shows that as the BD-RIS moves closer to the target, the CRB improves. This trend holds across all four schemes, with the proposed algorithm achieving the largest CRB improvement as the BD-RIS moves closer to the target. 
Unlike Benchmark Scheme II, the BD-RIS design enables high positioning accuracy even when the target is farther away. These results offer guidance for deploying BD-RIS in localization tasks.

\section{Conclusion}
In this letter, a BD-RIS-aided target localization was proposed. Applying BD-RIS to explore CRB performance of the target sensing is a challenging and difficult task. Specifically, it is difficult to satisfy the orthogonal unitary constraint of the BD-RIS scattering matrix while optimizing the CRB.  We proposed an efficient algorithm based on Riemannian complex space to construct a suitable gradient.  Simulation results demonstrated that the proposed scheme outperforms traditional RIS and benchmarks in sensing performance. Future work will explore multi-target and distributed BD-RIS scenarios for broader applicability.

\appendices
\section{Proof of Lemma 1} \label{Appendix A}
According to \cite{qiutidu}, we first give the standard definition to the derivative of $g({\bf{\Phi }})$ given in (16) with respect to ${\bf{\Phi }}$
\begin{equation}
        {\mathop{\rm d}\nolimits} g( {\bf{\Phi }} ) = \tfrac{{\partial g( {\bf{\Phi }} )}}{{\partial {\bf{\Phi }}}}{\mathop{\rm d}\nolimits} {\bf{\Phi }} + \tfrac{{\partial g( {\bf{\Phi }})}}{{\partial {{\bf{\Phi }}^*}}}{\mathop{\rm d}\nolimits} {{\bf{\Phi }}^*}.
\end{equation}
For convenience, let  ${g _1} ( {\bf{\Phi }})= {\mathop{\rm tr}\nolimits} ( {{\bf{\dot h}}( {\bf{\Phi }} ){}{{{\bf{\dot h}}}^H}( {\bf{\Phi }} )} )$, ${g _2}( {\bf{\Phi }} ) = {{{{| {{\mathop{\rm tr}} ( {\bf{A}})} |}^2}} {/
 {\vphantom {{{{| {{\mathop{\rm tr}\nolimits} ( {\bf{A}} )} |}^2}} {{\mathop{\rm tr}\nolimits} ( {\bf{B}})}}}
 \kern-\nulldelimiterspace} {{\mathop{\rm tr}\nolimits} ( {\bf{B}} )}}$, where ${\bf{A}} = {\bf{h}}( {\bf{\Phi }}){}{{\bf{\dot h}}^H}( {\bf{\Phi }} )$, ${\bf{B}} = {\bf{h}}( {\bf{\Phi }}){{\bf{h}}^H}( {\bf{\Phi }})$. Thus  (16)  can be equivalent to $g ( {\bf{\Phi }} ) = {g _1}( {\bf{\Phi }})- {g _2}( {\bf{\Phi }})$.

Based on the rules of complex matrix derivatives as detailed in \cite{qiutidu},  the complex differential of ${f _1}$ is
\begin{equation}\label{18}
\begin{array}{*{20}{l}}
{{\rm{d}}{g_1}( {\bf{\Phi }} )}\\ = {\rm{tr}}\{ {{\bf{G\Phi }\bf{\dot a}_{\rm RIS}}{\bf{\dot a}_{\rm RIS}}^H{\rm{d}}( {{{\bf{\Phi }}^H}} ){\bf{G}}_{}^H \!+\! {\bf{G}}{\rm{d}}( {\bf{\Phi }} ){{\bf{\dot a}_{\rm RIS}} {\bf{\dot a}_{\rm RIS}}^H}{{\bf{\Phi }}^H}{\bf{G}}_{}^H} \}\\
{ = {\rm{tr}}[ {{{\bf{\Lambda }}_1}{\rm{d}}( {\bf{\Phi }} ) + {{\bf{\Lambda }}_2}{\rm{d}}( {{{\bf{\Phi }}^*}})} ],}
\end{array}
\end{equation}
where ${{{\bf{\Lambda }}_1} = {{{\bf{\dot a}_{\rm RIS}} {\bf{\dot a}_{\rm RIS}}^H}}{{\bf{\Phi }}^H}{\bf{G}}}$ and ${{{\bf{\Lambda }}_2} = {{( {{\bf{G}}_{}^H{\bf{G\Phi {{\bf{\dot a}_{\rm RIS}} {\bf{\dot a}_{\rm RIS}}^H}}}} )}^T}}$.

For ${g _2}( {\bf{\Phi }})$, the differential can be obtained as follows
\begin{equation}\label{18}
        \begin{array}{l}
                {\rm{d}}{g _2} ( {\bf{\Phi }} )=\! \frac{{{\rm{d}}[ {{\mathop{\rm tr}\nolimits} ( {\bf{A}}){{( {{\mathop{\rm tr}\nolimits} ( {\bf{A}} )} )}^*}} ]{\mathop{\rm tr}\nolimits}( {\bf{B}}) \!-\! {\mathop{\rm tr}\nolimits} ( {\bf{A}} ){{( {{\mathop{\rm tr}\nolimits}( {\bf{A}})})}^*}{\rm{d}}[ {{\mathop{\rm tr}\nolimits}( {\bf{B}} )} ]}}{{{{[ {{\mathop{\rm tr}\nolimits}( {\bf{B}})}]}^2}}}.
        \end{array}
\end{equation}
Let ${g _3} \left( {\bf{\Phi }} \right)=  {{\mathop{\rm tr}\nolimits} \left( {\bf{A}} \right){{\left( {{\mathop{\rm tr}\nolimits} \left( {\bf{A}} \right)} \right)}^*}} $, then the differential is
\begin{equation}\label{18}
\begin{array}{*{20}{l}}
{{\rm{d}}{g_3} \left( {\bf{\Phi }} \right)= {\rm{tr}}\left\{ {{\bf{G}}{\rm{d}}\left( {\bf{\Phi }} \right){\bf{\Omega }}{{\bf{\Phi }}^H}{\bf{G}}_{}^H + {\bf{G\Phi \Omega }}{\rm{d}}\left( {{{\bf{\Phi }}^H}} \right){\bf{G}}_{}^H} \right\}}\\
{\qquad \quad \ \ \! = {\rm{tr}}\left[ {{{\bf{C}}_1}{\rm{d}}\left( {\bf{\Phi }} \right) + {{\bf{C}}_2}{\rm{d}}\left( {{{\bf{\Phi }}^*}} \right)} \right]},
\end{array}
\end{equation}
where ${{\bf{C}}_1} = {\bf{\Omega }}{{\bf{\Phi }}^H}{\bf{G}}_{}^H{\bf{G}}$ and ${{\bf{C}}_2} = {( {{\bf{G}}_{}^H{\bf{G\Phi \Omega }}} )^T}$.

Then the differential of ${\mathop{\rm tr}\nolimits}( {\bf{B}} )$  can be expressed as
\begin{equation}
\begin{array}{*{20}{l}}
{{\rm{d}}[ {{\rm{tr}}( {\bf{B}})}]}
{ = {\rm{tr}}[ {{{\bf{D }}_1}{\rm{d}}( {\bf{\Phi }} ) + {{\bf{D }}_2}{\rm{d}}( {{{\bf{\Phi }}^*}})}],}
\end{array}
\end{equation}
where ${{\bf{D }}_1}\!=\!{{{\bf{ a}_{\rm RIS}} {{\bf a}_{\rm RIS}^H}}}{{\bf{\Phi }}^H}{\bf{G}}_{}^H{\bf{G}}$, ${{{\bf{D }}_2}\!=\!{{( {{\bf{G}}_{}^H{\bf{G\Phi }}{{\bf{ a}_{\rm RIS}} {{\bf a}_{\rm RIS}^H}}} )}^T}}$.

Thus, 
\begin{equation}
\begin{array}{l}
{\rm{d}}g( {\bf{\Phi }}) = {\rm{tr}}[ {{{\bf{\Lambda }}_1} - \frac{{{{\bf{L}}_1}}}{{{\rm{tr}}( {\bf{B}})}} + \frac{{r( {\bf{A}}){{( {{\rm{tr}}( {\bf{A}} )} )}^*}{{\bf{\Gamma }}_1}}}{{{{[ {{\rm{tr}}( {\bf{B}})} ]}^2}}}}]{\rm{d}}( {\bf{\Phi }})\\
\qquad \quad + {\rm{tr}}[ {{{\bf{\Lambda }}_2} - \frac{{{{\bf{L}}_2}}}{{{\rm{tr}}( {\bf{B}} )}} + \frac{{{\rm{tr}}( {\bf{A}} ){{( {{\rm{tr}}( {\bf{A}} )} )}^*}{{\bf{\Gamma }}_2}}}{{{{[ {{\rm{tr}}( {\bf{B}})}]}^2}}}} ]{\rm{d}}( {{{\bf{\Phi }}^*}}).
\end{array}
\end{equation}
According to the differential forms presented in Table 3.2 of \cite{RSD}, the Euclidean gradient of $g( {\bf{\Phi }} )$ with respect to $\bf{\Phi}^*$ can be obtained accordingly
\begin{equation}\label{18}
\begin{array}{l}
\nabla _{\rm Euc}^{( t )}g ( {\bf{\Phi }}) \buildrel \Delta \over = \frac{{\partial g( {\bf{\Phi }} )}}{{\partial {{\bf{\Phi }}^*}}}  = {[ {{{\bf{\Lambda }}_2} - \frac{{{{\bf{C}}_2}}}{{{\mathop{\rm tr}\nolimits} ( {\bf{B}} )}} + \frac{{{\mathop{\rm tr}\nolimits} ( {\bf{A}} ){{( {{\mathop{\rm tr}\nolimits} ( {\bf{A}})} )}^*}{{\bf{D }}_2}}}{{{{[ {{\mathop{\rm tr}\nolimits} ( {\bf{B}})} ]}^2}}}} ]^T}.
\end{array}
\end{equation}

\end{document}